\documentclass[aps,floats]{revtex4}
\usepackage{amsmath,amssymb}
\usepackage{graphicx,epsfig}
\usepackage[greek,english]{babel}
\usepackage{bbold}

\begin{document}
\bibliographystyle {plain}

\pdfoutput=1
\def\oppropto{\mathop{\propto}} 
\def\opsimeq{\mathop{\simeq}}
\def\opoverderline{\mathop{\overline}}
\def\operarrow{\mathop{\longrightarrow}}
\def\opsim{\mathop{\sim}}

\def\fig#1#2{\includegraphics[height=#1]{#2}}
\def\figx#1#2{\includegraphics[width=#1]{#2}}


\title{ Large Deviations for dynamical fluctuations of Open Markov processes, \\
with application to random cascades on trees
} 


\author{ C\'ecile Monthus }
 \affiliation{Institut de Physique Th\'{e}orique, 
Universit\'e Paris Saclay, CNRS, CEA,
91191 Gif-sur-Yvette, France}

\begin{abstract}

The large deviations at 'Level 2.5 in time' for time-dependent ensemble-empirical-observables, introduced by C. Maes, K. Netocny and B. Wynants [Markov Proc. Rel. Fields. 14, 445 (2008)] for the case of $N$ independent Markov jump processes, are extended to the case of open Markov processes with reservoirs : explicit formulas are given for the joint probability of empirical occupation numbers and empirical flows, both for discrete-time dynamics and for continuous-time jump dynamics, with possibly time-dependent dynamical rules and/or time-dependent driving of the reservoirs. This general formalism is then applied to random cascades on trees, where particles are injected at the root via a 'source reservoir', while the particles are removed at the leaves of the last generation of the tree via 'sink reservoirs'.

\end{abstract}

\maketitle


\section{ Introduction }

While the standard classification of Large Deviations (see the reviews \cite{oono,review_touchette} and references therein)
involves three Levels, with Level 1 for empirical observables, 
Level 2 for the empirical measure, and Level 3 for the empirical process,
the new 'Level 2.5' concerning the joint distribution of the time-empirical-densities and the time-empirical-flows 
over some large time interval $T$
has turned out to be the appropriate Level for Markov processes :
 the rate function for the Large Deviations with respect to $T$
can be then written as a very general explicit local-in-space functional
 for a single Markov chain \cite{fortelle_thesis,fortelle_chain,review_touchette} (see the reminder in Appendix \ref{app_singlemarkovchain}),
for a single Markov jump process \cite{fortelle_thesis,fortelle_jump,maes_canonical,maes_onandbeyond,wynants_thesis,chetrite_formal,BFG1,BFG2}
 (see the reminder in Appendix \ref{app_singlemarkovjump})
and for a single diffusion process \cite{wynants_thesis,maes_diffusion,chetrite_formal,engel}.
In addition, this  'Level 2.5' formulation allows to reconstruct any time-additive observable
of the dynamical trajectory via its decomposition in terms of the empirical densities and of the empirical flows.
As a consequence, this 'Level 2.5' framework gives an interesting alternative point of view
with respect to the studies focusing on the generating functions of time-additive observables 
via deformed Markov operators that have attracted a lot of interest recently
in various models
 \cite{derrida-lecture,lecomte_chaotic,lecomte_thermo,lecomte_formalism,
lecomte_glass,kristina1,kristina2,chetrite_canonical,chetrite_conditioned,lazarescu_companion,lazarescu_generic,touchette_langevin,derrida-conditioned,bertin-conditioned}.

In Ref. \cite{maes_onandbeyond}, another step further has been introduced
that could be called  'Level 2.5 in time' :
this name will become clear below with Eq. \ref{ratedyna}, where the rate function contains
an integral over time of time-dependent-observables.
Instead of focusing on a single Markov process as above, one considers
 a large number $N$ of independent Markov Jump processes with trajectories $x^{(k)}(t)$
with $k=1,2,..,N$
constructed from the same transition rates $w(y,x)$ and from the same initial condition $P_{ini}(x)$.
The idea is to analyze the dynamical fluctuations
of the ensemble-empirical-density at configuration $x$ at time $t$
\begin{eqnarray}
\rho_t(x) = \frac{1}{N} \sum_{k=1}^N \delta \left( x^{(k)}(t) -x \right)
\label{ensembleempirical}
\end{eqnarray}
and of the ensemble-empirical-transition-rates $k_t(y,x)  $
defined as the ratios between the empirical number of jumps from $x$ to $y$ at time $t$
and the number of particles that are present at $x$ at time $t$
\begin{eqnarray}
k_t(y,x) = \frac{ \displaystyle \sum_{k=1}^N \delta \left( x^{(k)}(t^+) -y \right) \delta \left( x^{(k)}(t) -x \right)}
{ \displaystyle  \sum_{k=1}^N \delta \left( x^{(k)}(t) -x \right)} 
\label{ensemblejump}
\end{eqnarray}
By consistency, these empirical observables should satisfy the empirical-master-equation 
\begin{eqnarray}
\partial_t \rho_{t}(x) = \sum_{y \ne x} \left( k_t(x,y) \rho_t(y)  - k_t(y,x)  \rho_{t}(x) \right)
\label{tconstraint}
\end{eqnarray}
The probability to observe the time-dependent ensemble-empirical-densities $\rho_{0 \leq t \leq T}(.)$ and 
the time-dependent ensemble-empirical-transition-rates $k_{0 \leq t \leq T}(.,.) $
then follows the large deviation form with respect to $N \to +\infty$
\begin{eqnarray}
 {\cal P} \left [ \rho_{0 \leq t \leq T} (.) , k_{0 \leq t \leq T}(.,.) \right] 
= e^{ - N {\cal I}  \left[ \rho_{0 \leq t \leq T} (.) , k_{0 \leq t \leq T}(.,.)   \right]    }
\label{lddyna}
\end{eqnarray}
with the explicit rate function (Eqs 4.1, 4.2 and 4.3 of \cite{maes_onandbeyond} written with our present notations)
\begin{eqnarray}
 {\cal I}  \left[ \rho_{0 \leq t \leq T} (.) , k_{0 \leq t \leq T}(.,.)    \right]  
 && 
=\sum_x \rho_0(x) \ln \left( \frac{\rho_0(x) }{ P_{ini}(x) } \right)
 \nonumber \\ &&
 +  \int_0^T dt \sum_{x} \sum_{y \ne x} 
\left[   k_t(y,x) \ln \left( \frac{ k_t(y,x)  }{  w(y,x)  } \right) -  k_t(y,x)   + w(y,x) \right]\rho_t(x)
\label{ratedyna}
\end{eqnarray}
The first term is a boundary contribution coming from the initial condition,
while the second term is an integral over the time-interval $T$
of a functional of the time-dependent-empirical-observables that is local both in space and time.
Moreover, the special case of stationary ensemble-empirical observables $\rho_t(x)=\rho(x)$ and $k_t(y,x)=k(y,x)$
\cite{maes_onandbeyond} allows to recover 
the standard 'Level 2.5' formula for time-empirical-observables recalled in Eq. \ref{level2.75masterk},
while the empirical dynamics of Eq. \ref{tconstraint} reduces to the standard stationarity constraint of Eq. \ref{contraintek}.

The 'Level 2.5 in time' result of Eq. \ref{ratedyna}
has been derived in \cite{maes_onandbeyond} for 
$N$ independent Markov Jump processes with trajectories $x^{(k)}(t)$
by the method of change of measure on the trajectories.
The goal of the present paper is to extend this 'Level 2.5 in time' approach to 
open Markov processes with reservoirs, where the total number $N$ of Markov processes is not fixed anymore,
so that the ensemble-empirical-density at configuration $x$ at time $t$ of Eq. \ref{ensembleempirical}
will be replaced by the ensemble-empirical-occupation-numbers 
\begin{eqnarray}
N_t(x) = \sum_{k} \delta \left( x^{(k)}(t) -x \right)
\label{ntx}
\end{eqnarray}
that should be large enough to consider the large deviations with respect to $N_t(x)$ for each $x$ and $t$.

The paper is organized as follows.
We first analyze the large deviations at  'Level 2.5 in time' for general open Markov processes
involving reservoirs 
with possibly time-dependent dynamical rules, both for discrete time dynamics in section \ref{sec_discrete}
and for continuous time dynamics in section \ref{sec_continuous}.
We then describe the application of this general approach to the specific case of
random cascades on trees between a 'source reservoir' at the root and 'sink reservoirs' on the leaves of the last generation,
again both for discrete time dynamics in section \ref{sec_discretetree}
and for continuous time dynamics in section \ref{sec_continuoustree}.
Our conclusions are summarized in \ref{sec_conclusion}.
The Appendix \ref{app_reminder} contains a brief reminder on the Large deviations 'at Level 2.5' for the time-empirical-observables of a single Markov chain and a single Markov jump process respectively, in order to make comparisons with equations of the text.


\section{ Discrete time Markov processes with reservoirs  } 

\label{sec_discrete}

At time $t$, the state of the system is defined by 
 the empirical numbers $N_t(x)$ of particles on each discrete position $x=1,..,\Omega$ of the system,
while the state of the reservoirs is defined by the numbers $N^{ext}_t(r)$ of particles on each reservoir
site $r=1,2,..,R$.
The reservoirs numbers $N^{ext}_t(r)$ are fixed by the external protocol and do not depend on the system variables, for instance they can be fixed to be time-independent $N^{ext}_t(r)=N^{ext}(r)$ or to follow some time-dependence.

\subsection{ Model and notations }

\label{sec_dyn}

At time $t$, the state of the system is defined by 
 the empirical numbers $N_t(x)$ of particles on each discrete position $x=1,..,\Omega$ of the system,
while the state of the reservoirs is defined by the numbers $N^{ext}_t(r)$ of particles on each reservoir
site $r=1,2,..,R$.
The reservoirs numbers $N^{ext}_t(r)$ are fixed by the external protocol and do not depend on the system variables, for instance they can be fixed to be time-independent $N^{ext}_t(r)=N^{ext}(r)$ or to follow some time-dependence.

The dynamics is defined as follows.
All particles can move independently.

Between $t$ and $(t+1)$, a particle that was on site $x$ of the system at time $t$ can  :

(i) either move towards a position $y=1,2,..,\Omega$ of the system with probabilities $W_t(y,x)$
(including $y=x$ corresponding to the case where the particle remains on site $x$ with probability $W_t(x,x)$).

(ii) or move towards one of the reservoirs $r=1,2,..,R$ with probabilities $W_t(r,x)$

These probabilities are normalized to unity for each $x$ and for each $t$
\begin{eqnarray}
1 = \sum_{y=1}^{\Omega} W_t(y,x) + \sum_{r=1}^R W_t(r,x) 
\label{normawoutofx}
\end{eqnarray}

Between $t$ and $(t+1)$, a particle on reservoir $r$ at time $t$ can 
 move towards one of the sites of the system $y$ with probabilities $W_t(y,r)$ 
or remain on the reservoir $r$ with probability $W_t(r,r)$ with the normalization of probabilities
\begin{eqnarray}
1 = \sum_{y=1}^{\Omega} W_t(y,r) + W_t(r,r) 
\label{normawoutofr}
\end{eqnarray}

\subsection{ Large deviation analysis of the empirical transition probabilities out of a given site $x$ at time $t$  }

The number $N_t(x)$ of particles on each position $x $ at time $t$
is assumed to be large enough 
in order to consider large deviation with respect to $N_t(x)$.

\subsubsection{ Direct analysis in terms of the multinomial distribution of the outgoing flows }

Among these $N_t(x)$ independent particles on site $x$ at time $t$,
one considers the empirical numbers $Q_t(y,x)$ of particles that jump towards $y =1,..,\Omega$, 
and the numbers $Q_t(r,x)$ of particles that jump towards the reservoirs $r=1,..,R$
 with the normalization
\begin{eqnarray}
N_t(x) =  \sum_{y=1}^{\Omega} Q_t(y,x) + \sum_{r=1}^R Q_t(r,x)
\label{ntxoutflows}
\end{eqnarray}
Since the dynamics for each independent particle is drawn with the transition probabilities of Eq. \ref{normawoutofx},
the joint distribution of the flows $Q_t(y,x) $ and $Q_t(r,x) $ is simply the multinomial distribution
\begin{eqnarray}
&&  P[ Q_t(.,x) \vert  N_t(x) ]     = 
\nonumber \\
&& \ \ \   \frac{ N_t(x) !}{ \displaystyle \prod_{y=1}^{\Omega}  Q_t(y,x)!  \prod_{r=1}^{R}  Q_t(r,x)!  }  
 \prod_{y=1}^{\Omega} [ W_t(y,x)]^{Q_t(y,x)} \prod_{r=1}^{R} [ W_t(r,x)]^{Q_t(r,x)}
\delta_{ N_t(x) ,  \sum_{y=1}^{\Omega} Q_t(y,x) + \sum_{r=1}^R Q_t(r,x) }
\label{multinomial}
\end{eqnarray}
Using the Stirling's approximation for the factorials $m! \simeq \sqrt{2 \pi m} \ m^m e^{-m}$,
one obtains that the probability of the empirical transition probabilities 
\begin{eqnarray}
K_t(y,x) = \frac{Q_t(y,x) }{N_t(x) }
\nonumber \\
K_t(r,x) = \frac{Q_t(r,x) }{N_t(x) }
\label{p1k}
\end{eqnarray}
normalized to
\begin{eqnarray}
1 =  \sum_{y=1}^{\Omega} K_t(y,x) + \sum_{r=1}^R K_t(r,x)
\label{normap1k}
\end{eqnarray}
follows the large deviation form with respect to $N_t(x)$
\begin{eqnarray}
 P[ K_t(.,x) \vert N_t(x)]  \opsimeq_{N_t(x) \to +\infty}  e^{ - N_t(x) I[ K_t(.,x) ]  }  \delta_{ 1,  \sum_{y=1}^{\Omega} K_t(y,x) + \sum_{r=1}^R K_t(r,x) }
\label{largedevntx}
\end{eqnarray}
where the rate function $I[ K_t(.,x) ]  $ corresponds to the relative entropy of the empirical transition probabilities $K_t(.,x)$
with respect to the true transition probabilities $W_t(.,x)$
\begin{eqnarray}
 I[ K_t(.,x) ] = \sum_{y =1}^{\Omega} K_t(y,x)   \ln \left( \frac{K_t(y,x)  }{  W_t(y,x) } \right)
 + \sum_{r=1}^R K_t(r,x)   \ln \left( \frac{K_t(r,x)  }{  W_t(r,x) } \right) 
\label{Kxsanov}
\end{eqnarray}
This type of result based on the application of the Stirling's approximation to the multinomial distribution of Eq. \ref{multinomial} goes back to Boltzmann \cite{ellis} but is usually called Sanov Theorem in the large deviation language
\cite{oono,review_touchette,ellis}.

\subsubsection{ Alternative analysis in terms of the generating function of the outgoing flows }

Equivalently, one can consider the generating function of the flows $Q_t(y,x) $ and $Q_t(r,x) $ out of $x$ at time $t$
\begin{eqnarray}
Z[ \nu_t(.,x) \vert N_t(x) ]    
&& \equiv \sum_{Q_t(.,x) } P[ Q_t(.,x)  ] \ \  e^{ \displaystyle \sum_{y=1}^{\Omega} \nu_t(y,x) Q_t(y,x) + \sum_{r=1}^R \nu_t(r,x) Q_t(r,x) }
\nonumber \\
&& \equiv \sum_{K_t(.,x) } P[ K_t(.,x)  ] \ \  e^{ N_t(x) \left[ \displaystyle \sum_{y=1}^{\Omega} \nu_t(y,x) K_t(y,x) + \sum_{r=1}^R \nu_t(r,x) K_t(r,x) \right] }
\label{genemultinomial}
\end{eqnarray}
whose explicit expression for $N_t(x)$ independent particles is directly
\begin{eqnarray}
Z[ \nu_t(.,x) \vert N_t(x) ]    
= \left[ \sum_{y=1}^{\Omega} W_t(y,x) e^{\nu_t(y,x)  } + \sum_{r=1}^R W_t(r,x)  e^{\nu_t(r,x)  } \right]^{N_t(x)}
= e^{ N_t(x) \Phi[ \nu_t(.,x)]   }
\label{genemultinomialres}
\end{eqnarray}
where
\begin{eqnarray}
\Phi[ \nu_t(.,x)] =
\ln  \left[ \sum_{y=1}^{\Omega} W_t(y,x) e^{\nu_t(y,x)  } + \sum_{r=1}^R W_t(r,x)  e^{\nu_t(r,x)  } \right]
\label{genemultinomialphi}
\end{eqnarray}
is the Legendre transform of the rate function of Eq. \ref{Kxsanov} submitted to the normalization of Eq. \ref{normap1k}
\begin{eqnarray}
\Phi[ \nu_t(.,x)] =  \max \limits_{ K_t(.,x) : 1=\sum_y K_t(y,x)+\sum_r K_t(r,x)  } \left[ \sum_{y=1}^{\Omega} \nu_t(y,x) K_t(y,x) + \sum_{r=1}^R \nu_t(r,x) K_t(r,x)  - I [K_t(.,x)   ] \right]
\label{genemultinomiallegendre}
\end{eqnarray}
as a consequence of the saddle-point evaluation for large $N_t(x)$ of Eq. \ref{genemultinomial}.
The reciprocal Legendre transform
\begin{eqnarray}
 I [K_t(.,x)  ] =  \max \limits_{ \nu_t(.,x)   } \left[ \sum_{y=1}^{\Omega} \nu_t(y,x) K_t(y,x) + \sum_{r=1}^R \nu_t(r,x) K_t(r,x)  - \Phi[ \nu_t(.,x)] \right]
\label{genemultinomiallegendrereci}
\end{eqnarray}
allows to recover the result of Eq. \ref{Kxsanov} by this Legendre transform of Eq. \ref{genemultinomialphi}.

\subsubsection{ Alternative analysis in terms of a change of measure }

A third derivation of Eq. \ref{Kxsanov} consists in considering a change of measure :
the empirical flows $Q_t(.,x)$ would be typical if one replaces the true transition probabilities $W_t(.,x)$
by the empirical transition probabilities $K_t(.,x)$ of Eq. \ref{p1k},
so that the corresponding modified multinomial distribution of Eq. \ref{multinomial} becomes
\begin{eqnarray}
&& P^{mod}[ Q_t(.,x) \vert N_t(x) ]   
\nonumber \\
&& \ \ \  =  \frac{ N_t(x) !}{ \displaystyle \prod_{y=1}^{\Omega}  Q_t(y,x)!  \prod_{r=1}^{R}  Q_t(r,x)!  }  
 \prod_{y=1}^{\Omega} [ K_t(y,x)]^{Q_t(y,x)} \prod_{r=1}^{R} [ K_t(r,x)]^{Q_t(r,x)}
\delta_{ N_t(x) ,  \sum_{y=1}^{\Omega} Q_t(y,x) + \sum_{r=1}^R Q_t(r,x) }
\label{multinomialbis}
\end{eqnarray}
As a consequence, the large-deviation form 
can be also obtained directly from the ratio
\begin{eqnarray}
&& \frac{ P[ Q_t(.,x) \vert N_t(x) ]}{ P^{mod}[ Q_t(.,x) \vert N_t(x) ]   }
 = 
 \prod_{y=1}^{\Omega} \left[ \frac{W_t(y,x)}{K_t(y,x)}\right]^{Q_t(y,x)} \prod_{r=1}^{R} \left[ \frac{W_t(r,x)}{K_t(r,x)} \right]^{Q_t(r,x)}
\delta_{ N_t(x) ,  \sum_{y=1}^{\Omega} Q_t(y,x) + \sum_{r=1}^R Q_t(r,x) }
\nonumber \\
&& =  e^{ - \displaystyle \sum_{y=1}^{\Omega}Q_t(y,x) \ln \left[ \frac{Q_t(y,x)}{W_t(y,x) N_t(x) }\right]
-  \sum_{r=1}^{R} Q_t(r,x) \ln \left[ \frac{Q_t(r,x)}{W_t(r,x) N_t(x) } \right] }
\delta_{ N_t(x) ,  \sum_{y=1}^{\Omega} Q_t(y,x) + \sum_{r=1}^R Q_t(r,x) }
\label{ratiomultinomialbis}
\end{eqnarray}
that corresponds to Eq. \ref{largedevntx} and \ref{Kxsanov} when translated into the 
empirical transition probabilities $K_t(.,x) = \frac{Q_t(.,x)}{N_t(x)}$.

\subsubsection{ Discussion of the hypothesis of large $N_t(x)$ }

\label{sec_hyp}

As stated after Eq. 6 in the Introduction, the main hypothesis is that $N_t(x)$
should be large enough for all $x$ and all $t$ in order to be able to write
the large deviation form 
 with respect to $N_t(x)$ for the outgoing flows out of site $x$ at time $t$  (Eq. \ref{largedevntx}).
It is interesting to compare how this hypothesis appears more precisely  in
the three arguments above : 

(i) in the first direct analysis, it appears via the Stirling's approximation
needed to go from the multinomial distribution of Eq. \ref{multinomial} to the rate function of Eq. \ref{Kxsanov}.

(ii) in the second alternative analysis, the generating function of Eq. \ref{genemultinomialres}
is actually valid for any finite $N_t(x)$, so that the condition of large $N_t(x)$
only appears when one wishes to perform the inverse Legendre transform via the saddle-point approximation
(Eqs \ref{genemultinomiallegendre} and \ref{genemultinomiallegendrereci}).

(iii) in the third alternative analysis, the hypothesis of large $N_t(x)$ is somewhat hidden 
in the fact that one should be able to define the empirical transition probabilities $K_t(.,x) = \frac{Q_t(.,x)}{N_t(x)}$
(for instance they would loose their meaning if some empirical occupations $N_t(x)$ can vanish),
and one should be able to consider that the comparison with the modified process of Eq. \ref{multinomialbis}
via the ratio of Eq. \ref{ratiomultinomialbis} is enough to obtain the correct result in the exponential.

As a final remark, we should stress that in the present paper,
in order to have a more straighforward physical interpretation,
we have chosen to interpret $N_t(x)$ as the empirical number of independent particles really present
at some position $x$ at time $t$. But the ensemble-empirical-framework introduced in 
Ref. \cite{maes_onandbeyond} is of course much more general,
since $x$ can label more general configuration spaces,
and the notion of ensemble can be interpreted in the traditional sense of statistical physics
as a very large number of virtual copies of the elementary dynamical system one is interested in.

\subsection{ Large deviation analysis of the empirical transition probabilities out of a given reservoir $r$ at time $t$  }

The same analysis for the empirical transition probabilities $K_t(x,r)$ out of the reservoir $r$ yields
the large deviation form with respect to $N^{ext}_t(r)$
\begin{eqnarray}
 P[ K_t(.,r) \vert N^{ext}_t(r) ]  \opsimeq_{N^{ext}_t(r) \to +\infty}  e^{ - N^{ext}_t(x) I[ K_t(.,r) ]  }  \delta_{ 1,  \sum_{y=1}^{\Omega} K_t(y,r) +  K_t(r,r) }
\label{largedevntr}
\end{eqnarray}
where the rate function $I[ K_t(.,r) ]  $ corresponds to the relative entropy of the empirical transition probabilities $K_t(.,r)$
with respect to the true transition probabilities $W_t(.,r)$
\begin{eqnarray}
 I[ K_t(.,r) ] = \sum_{y =1}^{\Omega} K_t(y,r)   \ln \left( \frac{K_t(y,r)  }{  W_t(y,r) } \right)
 +  K_t(r,r)   \ln \left( \frac{K_t(r,r)  }{  W_t(r,r) } \right) 
\label{Kxsanor}
\end{eqnarray}

\subsection{ Analysis of the state of the system at time $t+1$  }

For the reservoirs, the new numbers $N^{ext}_{t+1}(r)   $ are imposed by the external protocol.

In the system, the empirical number $N_{t+1}(x)$ of particles on site $x$ at time $(t+1)$
is completely determined by the empirical flows $Q_t(x,.)$ of particles towards $x$ between $t$ and $t+1$ 
\begin{eqnarray}
N_{t+1}(x) =\sum_{y=1}^{\Omega} Q_t(x,y) + \sum_{r=1}^R Q_t(x,r) 
= \sum_{y=1}^{\Omega} K_t(x,y)N_t(y) + \sum_{r=1}^R K_t(x,r) N_{t}^{ext}(r)
\label{nnextingoingflows}
\end{eqnarray}
Note the difference with $N_t(x)$ of Eq. \ref{ntxoutflows}
involving the flows $Q_t(.,x)$ out of $x$.

\subsection{ Probability of an empirical history of the occupation numbers $N_{t}(.)$ and flows $Q_{t}(.,.)$ }

Putting everything together, one obtains that
once the empirical occupation numbers $N_t(.)$ at time $t$ are known,
the joint probability of the empirical flows $Q_t(.,.)$ between $t$ and $(t+1)$ 
and of the empirical occupation numbers $N_{t+1}(.)$ at time $(t+1)$ 
reads 
\begin{eqnarray}
&&  P[ N_{t+1}(.) ;Q_t(.,.) \vert N_t(.)]  \simeq 
\nonumber \\
&& 
e^{ - \displaystyle 
\sum_{x =1}^{\Omega} \left[ \sum_{y =1}^{\Omega} Q_t(y,x)   \ln \left( \frac{Q_t(y,x)  }{  W_t(y,x) N_t(x)} \right)
 + \sum_{r=1}^R Q_t(r,x)   \ln \left( \frac{Q_t(r,x)  }{  W_t(r,x) N_t(x)} \right) \right] }
\prod_{x=1}^{\Omega} \delta_{N_{t}(x) , \sum_{y =1}^{\Omega}  Q_t(y,x)  + \sum_{r=1}^R Q_t(r,x)  }
\nonumber \\
&& e^{ - \displaystyle \sum_{r=1}^R \left[ \sum_{y=1}^{\Omega} Q_t(y,r)   \ln \left( \frac{Q_t(y,r)  }{  W_t(y,r) N^{ext}_t(r)} \right)
 + Q_t(r,r)   \ln \left( \frac{Q_t(r,r)  }{  W_t(r,r)N^{ext}_t(r) } \right) \right] }
\prod_{r=1}^{R} \delta_{N^{ext}_{t}(r) , \sum_{y =1}^{\Omega}  Q_t(y,r)  + Q_t(r,r)  }
\nonumber \\
&&  \prod_{x=1}^{\Omega} \delta_{N_{t+1}(x) , \sum_{y =1}^{\Omega}  Q_t(x,y)  + \sum_{r=1}^R Q_t(x,r)  }
\label{KNjoint}
\end{eqnarray}
where the second line corresponds to the statistics of the outgoing flows $Q_t(.,x)=K_t(.,x) N_t(x)$ out of the system sites $x$  (Eqs \ref{largedevntx} and \ref{Kxsanov}),
where the third line corresponds to the statistics of the outgoing flows $Q_t(.,r)=K_t(.,r) N^{ext}_t(r)$ out of the 
reservoirs $r$  (Eqs \ref{largedevntr} and \ref{Kxsanor}),
and where the fourth line corresponds to the empirical dynamics of Eq. \ref{nnextingoingflows}.

From this elementary Markov propagator, one obtains via iteration 
the joint probability of the empirical flows $Q_{0 \leq t \leq T-1}(.,.)$
and of the empirical occupation numbers $N_{1 \leq t \leq T}(.)$ 
given the empirical occupation numbers $N_{t=0}(.)$ at time $t=0$
\begin{eqnarray}
&& {\cal P} [ N_{1 \leq t \leq T}(.) ; Q_{0 \leq t \leq T-1}(.,.) \vert N_0(.)]
\nonumber \\
&&  = P[ N_{T}(.) ;Q_{T-1}(.,.) \vert N_{T-1}(.)]  P[ N_{T-1}(.) ;Q_{T-2}(.,.) \vert N_{T-2}(.)] ...
... P[ N_2(.) ;Q_1(.,.) \vert N_1(.)]  
 P[ N_1(.) ;Q_0(.,.) \vert N_0(.)] =
\nonumber \\
&&  
e^{ - \displaystyle 
\sum_{t=0}^{T-1} \sum_{x =1}^{\Omega} \left[ \sum_{y =1}^{\Omega} Q_t(y,x)   \ln \left( \frac{Q_t(y,x)  }{  W_t(y,x) N_t(x)} \right)
 + \sum_{r=1}^R Q_t(r,x)   \ln \left( \frac{Q_t(r,x)  }{  W_t(r,x) N_t(x)} \right) \right] }
\prod_{t=0}^{T-1} \prod_{x=1}^{\Omega} \delta_{N_{t}(x) , \sum_{y =1}^{\Omega}  Q_t(y,x)  + \sum_{r=1}^R Q_t(r,x)  }
\nonumber \\
&& e^{ - \displaystyle 
\sum_{t=0}^{T-1}\sum_{r=1}^R \left[ \sum_{y=1}^{\Omega} Q_t(y,r)   \ln \left( \frac{Q_t(y,r)  }{  W_t(y,r) N^{ext}_t(r)} \right)
 + Q_t(r,r)   \ln \left( \frac{Q_t(r,r)  }{  W_t(r,r)N^{ext}_t(r) } \right) \right] }
\prod_{t=0}^{T-1} \prod_{r=1}^{R} \delta_{N^{ext}_{t}(r) , \sum_{y =1}^{\Omega}  Q_t(y,r)  + Q_t(r,r)  }
\nonumber \\
&&  \prod_{t=0}^{T-1} \prod_{x=1}^{\Omega} \delta_{N_{t+1}(x) , \sum_{y =1}^{\Omega}  Q_t(x,y)  + \sum_{r=1}^R Q_t(x,r)  }
\label{ptrajmarkovchain}
\end{eqnarray}
This result concerning the probability of an empirical history of the occupation numbers $N_{t}(.)$ and flows $Q_{t}(.,.)$
is the most general formula of the present paper, and the remainder of the paper 
will be devoted to the analysis of various specific cases.

\subsection{Application to time-independent dynamics during a large time interval $T \to +\infty$ }

When the dynamics corresponds to time-independent transition probabilities $W_t(.,.)=W(.,.)$
and time-independent reservoirs occupation numbers $  N_t^{ext}(r) = N^{ext}(r)$ ,
it is natural to consider the probability to observe some stationary occupation numbers $N_t(.)=N(.)$
and flows $Q_t(.,.)=Q(.,.)$ during a large time interval $T \to +\infty$
(so that the initial condition is not important anymore). Eq. \ref{ptrajmarkovchain}
then yields the following large deviation form with respect to $T \to +\infty$
\begin{eqnarray}
&& {\cal P}^{statio} [ N(.) ; Q(.,.) ] \opsimeq_{T \to +\infty} 
\nonumber \\
&&  
e^{ - T \displaystyle 
 \sum_{x =1}^{\Omega} \left[ \sum_{y =1}^{\Omega} Q(y,x)   \ln \left( \frac{Q(y,x)  }{  W(y,x) N(x)} \right)
 + \sum_{r=1}^R Q(r,x)   \ln \left( \frac{Q(r,x)  }{  W(r,x) N(x)} \right) \right] }
 \prod_{x=1}^{\Omega} \delta_{N(x) , \sum_{y =1}^{\Omega}  Q(y,x)  + \sum_{r=1}^R Q(r,x)  }
\nonumber \\
&& e^{ - T \displaystyle 
\sum_{r=1}^R \left[ \sum_{y=1}^{\Omega} Q(y,r)   \ln \left( \frac{Q(y,r)  }{  W(y,r) N^{ext}(r)} \right)
 + Q(r,r)   \ln \left( \frac{Q(r,r)  }{  W(r,r)N^{ext}(r) } \right) \right] }
 \prod_{r=1}^{R} \delta_{N^{ext}(r) , \sum_{y =1}^{\Omega}  Q(y,r)  + Q(r,r)  }
\nonumber \\
&& \prod_{x=1}^{\Omega} \delta_{N(x) , \sum_{y =1}^{\Omega}  Q(x,y)  + \sum_{r=1}^R Q(x,r)  }
\label{pstatiomarkovchain}
\end{eqnarray}
Within the present derivation, this result inherits from Eq. \ref{ptrajmarkovchain}
the condition that the occupation numbers $N(x)$ should be large enough for all $x$,
since this is the general hypothesis of the present paper, as stressed in section \ref{sec_hyp}.

As already mentioned after Eq. \ref{ratedyna}, the case of stationary ensemble-empirical observables
is related to the case of time-empirical observables \cite{maes_onandbeyond}.
For our present case, this means that
the formula for $ {\cal P}^{statio} [ N(.) ; Q(.,.) ] $ also describes the probability ${\cal P}^{te} [ N^{te}(.) ; Q^{te}(.,.) ]  $ to observe the time-empirical observables
\begin{eqnarray}
N^{te} (x) \equiv \frac{1}{T} \sum_{t=1}^T N_t(x)
\nonumber \\
Q^{te} (.,.) \equiv \frac{1}{T} \sum_{t=0}^{T-1} Q_t(.,.)
\label{timeempirical}
\end{eqnarray}
since the easiest way to realize these time-empirical values in the dynamical formula of Eq. \ref{ptrajmarkovchain}
corresponds to the dynamical histories that remain stationary $(N_t(x)=N^{te} (x) ;Q_t(.,.) = Q^{te} (.,.) ) $ on the time-interval $[0,T]$ \cite{maes_onandbeyond}
\begin{eqnarray}
{\cal P}^{te} [ N^{te}(.) ; Q^{te}(.,.) ] ={\cal P}^{statio} [ N^{te}(.) ; Q^{te}(.,.) ] 
\label{probatimeempirical}
\end{eqnarray}

As a consequence, Eq. \ref{pstatiomarkovchain}
can be considered as the generalization in the presence of reservoirs
to the 'Level 2.5' rate function of Eq. \ref{alevel2.5chain} with the constraint of Eq. \ref{Cstatio}
concerning time-empirical observables for a single stationary Markov chain
\cite{fortelle_thesis,fortelle_chain,review_touchette} (for more details see the reminder in Appendix \ref{app_singlemarkovchain}).

\subsection{Application to periodic dynamics   }

Since there has been recent studies on the large deviation properties
of periodically driven systems \cite{singh,bertini,barato17,barato18},
it is interesting to consider the case
where the dynamical rules alternate periodically between transition probabilities $W_{2t}(.,.)=W_{e}(.,.)$
and reservoirs occupation numbers $  N_{2t}^{ext}(r) = N_e^{ext}(r)$ at even times $(2t)$,
and transition probabilities $W_{2t+1}(.,.)=W_{o}(.,.)$
and reservoirs occupation numbers $  N_{2t+1}^{ext}(r) = N_o^{ext}(r)$ at odd times $(2t+1)$.
It is then natural to consider the probability to observe some alternating occupation numbers $N_{2t}(.)=N_e(.)$
and $N_{2t+1}(.)=N_o(.)$
and alternating flows $Q_{2t}(.,.)=Q_e(.,.)$ and  $Q_{2t+1}(.,.)=Q_o(.,.)$ during a large time interval $T \to +\infty$
(so that the initial condition is not important anymore) : Eq. \ref{ptrajmarkovchain}
then yields the following large deviation form with respect to $T$
\begin{eqnarray}
&& {\cal P}^{periodic} [ N_e(.), N_o ; Q_e(.,.) ,Q_o(.,.) ] \opsimeq_{T \to +\infty} 
\nonumber \\
&&  
e^{ - \frac{T}{2} \displaystyle 
 \sum_{x =1}^{\Omega} \left[ \sum_{y =1}^{\Omega} Q_e(y,x)   \ln \left( \frac{Q_e(y,x)  }{  W_e(y,x) N_e(x)} \right)
 + \sum_{r=1}^R Q_e(r,x)   \ln \left( \frac{Q_e(r,x)  }{  W_e(r,x) N_e(x)} \right) \right] }
 \prod_{x=1}^{\Omega} \delta_{N_e(x) , \sum_{y =1}^{\Omega}  Q_e(y,x)  + \sum_{r=1}^R Q_e(r,x)  }
\nonumber \\
&&  
e^{ - \frac{T}{2} \displaystyle 
 \sum_{x =1}^{\Omega} \left[ \sum_{y =1}^{\Omega} Q_o(y,x)   \ln \left( \frac{Q_o(y,x)  }{  W_o(y,x) N_o(x)} \right)
 + \sum_{r=1}^R Q_o(r,x)   \ln \left( \frac{Q_o(r,x)  }{  W_o(r,x) N_o(x)} \right) \right] }
 \prod_{x=1}^{\Omega} \delta_{N_o(x) , \sum_{y =1}^{\Omega}  Q_o(y,x)  + \sum_{r=1}^R Q_o(r,x)  }
\nonumber \\
&& e^{ - \frac{T}{2}  \displaystyle 
\sum_{r=1}^R \left[ \sum_{y=1}^{\Omega} Q_e(y,r)   \ln \left( \frac{Q_e(y,r)  }{  W_e(y,r) N^{ext}_e(r)} \right)
 + Q_e(r,r)   \ln \left( \frac{Q_e(r,r)  }{  W_e(r,r)N_e^{ext}(r) } \right) \right] }
 \prod_{r=1}^{R} \delta_{N_e^{ext}(r) , \sum_{y =1}^{\Omega}  Q_e(y,r)  + Q_e(r,r)  }
\nonumber \\
&& e^{ - \frac{T}{2}  \displaystyle 
\sum_{r=1}^R \left[ \sum_{y=1}^{\Omega} Q_o(y,r)   \ln \left( \frac{Q_o(y,r)  }{  W_o(y,r) N^{ext}_o(r)} \right)
 + Q_o(r,r)   \ln \left( \frac{Q_o(r,r)  }{  W_o(r,r)N_o^{ext}(r) } \right) \right] }
 \prod_{r=1}^{R} \delta_{N_o^{ext}(r) , \sum_{y =1}^{\Omega}  Q_o(y,r)  + Q_o(r,r)  }
\nonumber \\
&& \prod_{x=1}^{\Omega} \delta_{N_o(x) , \sum_{y =1}^{\Omega}  Q_e(x,y)  + \sum_{r=1}^R Q_e(x,r)  }
\ \ \ \prod_{x=1}^{\Omega} \delta_{N_e(x) , \sum_{y =1}^{\Omega}  Q_o(x,y)  + \sum_{r=1}^R Q_o(x,r)  }
\label{periodicmarkovchain}
\end{eqnarray}
This formula is somewhat lengthy but the various contributions have a very clear physical meaning :
the second line corresponds to the statistics of the outgoing flows $Q_e(.,x)$ from sites $x$ of the system at even times,
the third line corresponds to the statistics of the outgoing flows $Q_o(.,x)$ from sites $x$ of the system at odd times,
the fourth line corresponds to the statistics of the outgoing flows $Q_e(.,r)$ from reservoirs $r$ at even times,
the fifth line corresponds to the statistics of the outgoing flows $Q_o(.,r)$ from reservoirs $r$  at odd times,
while the last line corresponds to the dynamical constraints on the incoming flows towards sites $x$ of the system
at even and odd times respectively.

\section{Continuous time Markov processes with reservoirs }

\label{sec_continuous}

\subsection{ Models and notations}

To see more clearly the similarities and differences with the previous section,
it is more convenient here to consider 
that the continuous-time jump Markov dynamics is obtained as
the limit of the discrete-time Markov dynamics described in \ref{sec_dyn},
where the time interval $(t,t+1)$ is replaced by $(t,t+dt)$ with infinitesimal $dt$.
The transition probabilities of Eq. \ref{normawoutofx} towards reservoirs $r$ and towards other sites $y \ne x$ of the system become of order $(dt)$ with corresponding transition rates $w_t(r,x)$ and $w_t(y,x)$
\begin{eqnarray}
W_t(r,x) \opsimeq w_t(r,x)dt
\nonumber \\
W_t(y,x) \opsimeq w_t(y,x)dt
\label{wratex}
\end{eqnarray}
while the conservation of probabilities of Eq. \ref{normawoutofx}
imposes that the probability to remain on site $x$ reads
\begin{eqnarray}
W_t(x,x) && \opsimeq 1- dt \left[ \sum_{y \ne x}  w_t(y,x)+ \sum_r  w_t(r,x) \right]
\label{wratexx}
\end{eqnarray}

Similarly, the transition probabilities of Eq. \ref{normawoutofr} out of the reservoir $r$ are of the form
\begin{eqnarray}
W_t(x,r) && \opsimeq w_t(x,r)dt
\nonumber \\
W_t(r,r) && \opsimeq 1- dt \left[  \sum_y  w_t(y,r) \right]
\label{wrater}
\end{eqnarray}

\subsection{ Large deviation for the empirical transition rates out of a given site $x$ at time $t$  }

The empirical transition probabilities $K_t(y\ne x,x)$ and $K_t(r,x)$ of Eq. \ref{p1k} normalized to Eq. \ref{normap1k}
will be also infinitesimal with transitions rates $k_t(y,x) $ and $k_t(r,x) $
\begin{eqnarray}
K_t(y,x) \opsimeq k_t(y,x)dt
\nonumber \\
K_t(r,x) \opsimeq k_t(y,x)dt
\label{ktxrates}
\end{eqnarray}
while the probability to remain on $x$ reads
\begin{eqnarray}
K_t(x,x) && \opsimeq 1- dt \left[ \sum_{y \ne x}  k_t(y,x)+ \sum_r  k_t(r,x) \right]
\label{ktxx}
\end{eqnarray}

Eq. \ref{largedevntx} then gives the following contribution in terms of the transition rates $k_t(.,x)$ out of $x$
\begin{eqnarray}
 P_{dt}[ k_t(.,x) \vert N_t(x)]   \to 
 e^{ - dt N_t(x) I[ k_t(.,x) ]  }  
\label{largedevntxdt}
\end{eqnarray}
where the rate function $I[ k_t(.,x) ]  $ obtained from Eq. \ref{Kxsanov} 
\begin{eqnarray}
 I[ k_t(.,x) ] = \sum_{y \ne x} \left[ k_t(y,x)   \ln \left( \frac{k_t(y,x)  }{  w_t(y,x) } \right)- k_t(y,x)+w_t(y,x)  
\right] 
 + \sum_{r=1}^R \left[ k_t(r,x)   \ln \left( \frac{k_t(r,x)  }{  w_t(r,x) } \right) 
- k_t(r,x)+w_t(r,x)  
\right]
\label{Kxsanovdt}
\end{eqnarray}
corresponds to Poisson processes.

\subsection{ Large deviation for the empirical transition rates out of a reservoir $r$ at time $t$  }

Similarly, Eq. \ref{largedevntr} gives the following contribution in terms of the transition rates $k_t(.,r)$ out of the reservoir $r$ at time $t$
\begin{eqnarray}
 P_{dt}[ k_t(.,r) \vert N^{ext}_t(r) ]  \to  e^{ - dt N^{ext}_t(x) I[ k_t(.,r) ]  }  
\label{largedevntrdt}
\end{eqnarray}
with the rate function $I[ k_t(.,r) ]  $ obtained from Eq. \ref{Kxsanor}
\begin{eqnarray}
 I[ k_t(.,r) ] = \sum_{y } \left[ k_t(y,r)   \ln \left( \frac{k_t(y,r)  }{  w_t(y,r) } \right)- k_t(y,r)+w_t(y,r)  
\right] 
\label{Kxsanordt}
\end{eqnarray}

\subsection{ Constraint from the empirical dynamics  }

The empirical dynamics of Eq. \ref{nnextingoingflows} becomes the differential equation
\begin{eqnarray}
\frac{d N_{t}(x)}{dt}  
=  \sum_{y \ne x}^{\Omega} \left( k_t(x,y)N_t(y) - k_t(y,x)N_t(x) \right) 
 + \sum_{r=1}^R \left( k_t(x,r) N_{t}^{ext}(r) - k_t(r,x) N_{t}(x) \right)
\label{empiricaldyna}
\end{eqnarray}

As a consequence, it is more convenient to replace the empirical transition rates $k_t(.,.)$ 
by the empirical flows per unit time
\begin{eqnarray}
q_t(x,y) && \equiv  k_t(x,y)N_t(y) 
\nonumber \\
q_t(x,r) && \equiv k_t(x,r) N_{t}^{ext}(r) 
\nonumber \\
q_t(r,x) && \equiv k_t(r,x) N_{t}(x) 
\label{empiricalflows}
\end{eqnarray}
to rewrite the dynamics of Eq. \ref{empiricaldyna} as
\begin{eqnarray}
\frac{d N_{t}(x)}{dt}  
=  \sum_{y \ne x}^{\Omega} \left( q_t(x,y) - q_t(y,x) \right) 
 + \sum_{r=1}^R \left( q_t(x,r)  - q_t(r,x) \right)
\label{empiricaldynaq}
\end{eqnarray}

\subsection{ Probability of an empirical history of the occupation numbers $N_{t}(.)$ and flows $q_{t}(.,.)$ }

Eq. \ref{ptrajmarkovchain} yields that the probability to observe the empirical occupation numbers $N_{0 \leq t \leq T}(.)$
and the empirical flows $ q_{0 \leq t \leq T}(.,.) $ reads
\begin{eqnarray}
  {\cal P} [ N_{0 \leq t \leq T}(.) ; q_{0 \leq t \leq T}(.,.)  \vert N_0(.)]
&& =
e^{ - \displaystyle 
\int_{0}^{T}dt  \sum_{x =1}^{\Omega} \sum_{y \ne x} \left[ q_t(y,x)   \ln \left( \frac{q_t(y,x)  }{  w_t(y,x) N_t(x)} \right)- q_t(y,x)+w_t(y,x)  N_t(x) \right] }
 \nonumber \\
&& e^{ - \displaystyle 
\int_{0}^{T}dt  \sum_{x =1}^{\Omega}  \sum_{r=1}^R \left[ q_t(r,x)   \ln \left( \frac{q_t(r,x)  }{  w_t(r,x) N_t(x)} \right) 
- q_t(r,x)+w_t(r,x) N_t(x) \right] }
\nonumber \\
&& e^{ - \displaystyle 
\int_{0}^{T}dt  \sum_{r=1}^R \sum_{y=1 }^{\Omega} \left[ q_t(y,r)   \ln \left( \frac{q_t(y,r)  }{  w_t(y,r) N_t^{ext}(r)} \right)- q_t(y,r)
+w_t(y,r) N_t^{ext}(r) \right] }
\label{ptrajmarkovjump}
\end{eqnarray}
provided the empirical dynamics of Eq. \ref{empiricaldynaq} is satisfied on the time interval $t \in [0,T]$ for all sites $x$ of the system.
This is the generalization in the presence of reservoirs of Eqs \ref{tconstraint} , \ref{lddyna} and \ref{ratedyna} 
quoted in the Introduction.

\subsection{Application to time-independent dynamics during a large time interval $T \to +\infty$ }

When the dynamics corresponds to time-independent transition rates $w_t(.,.)=w(.,.)$
and time-independent reservoirs occupation numbers $  N_t^{ext}(r) = N^{ext}(r)$,
one obtains that the probability to observe some stationary occupation numbers $N_t(.)=N(.)$
and flows $q_t(.,.)=q(.,.)$ during a large time interval $T \to +\infty$
follows the large deviation form with respect to $T$
\begin{eqnarray}
 {\cal P}^{statio} [ N(.) ; q(.,.) ] \opsimeq_{T \to +\infty} 
&&  
e^{ - \displaystyle 
T  \sum_{x =1}^{\Omega} \sum_{y \ne x} \left[ q(y,x)   \ln \left( \frac{q(y,x)  }{  w(y,x) N(x)} \right)- q(y,x)+w(y,x)  N(x) \right] }
 \nonumber \\
&& e^{ - \displaystyle 
T \sum_{x =1}^{\Omega}  \sum_{r=1}^R \left[ q(r,x)   \ln \left( \frac{q(r,x)  }{  w(r,x) N(x)} \right) 
- q(r,x)+w(r,x) N(x) \right] }
\nonumber \\
&& e^{ - \displaystyle 
T  \sum_{r=1}^R \sum_{y=1 }^{\Omega} \left[ q(y,r)   \ln \left( \frac{q(y,r)  }{  w(y,r) N^{ext}(r)} \right)- q(y,r)
+w(y,r) N_t^{ext}(r) \right] }
\nonumber \\
&&
\prod_{x=1}^{\Omega} \delta \left(  \sum_{y \ne x}^{\Omega} \left( q(x,y) - q(y,x) \right) + \sum_{r=1}^R \left( q(x,r)  - q(r,x) \right) \right)
\label{pstatiomarkovjump}
\end{eqnarray}
where the last line ensures that the empirical stationary dynamics $\frac{d N_{t}(x)}{dt}  
=  0$ of Eq. \ref{empiricaldynaq}
 is satisfied for all sites $x$ of the system.

As already discussed in the discrete case in Eqs \ref{timeempirical} and \ref{probatimeempirical},
this formula for $ {\cal P}^{statio} [ N(.) ; q(.,.) ] $ also describes the probability ${\cal P}^{te} [ N^{te}(.) ; q^{te}(.,.) ]  $ to observe the time-empirical observables \cite{maes_onandbeyond}
\begin{eqnarray}
N^{te} (x) \equiv \frac{1}{T} \int_0^T dt N_t(x)
\nonumber \\
q^{te} (.,.) \equiv \frac{1}{T} \int_0^T dt q_t(.,.)
\label{timeempiricalbis}
\end{eqnarray}
since the easiest way to realize these time-empirical values in the dynamical formula of Eq. \ref{ptrajmarkovjump}
corresponds to the dynamical histories that remain stationary $(N_t(x)=N^{te} (x) ;q_t(.,.) = q^{te} (.,.) ) $ on the time-interval $[0,T]$ \cite{maes_onandbeyond}
\begin{eqnarray}
{\cal P}^{te} [ N^{te}(.) ; q^{te}(.,.) ] ={\cal P}^{statio} [ N^{te}(.) ; q^{te}(.,.) ] 
\label{probatimeempiricalbis}
\end{eqnarray}
So the formula \ref{pstatiomarkovjump} can be considered as the generalization in the presence of reservoirs
to the 'Level 2.5' rate function of Eq. \ref{level2.75master} with the constraint of Eq. \ref{contrainteq}
concerning the time-empirical observables of a single stationary Markov jump process \cite{fortelle_thesis,fortelle_jump,maes_canonical,maes_onandbeyond,wynants_thesis,chetrite_formal,BFG1,BFG2}.

\section{ Application to discrete time Markovian cascades on trees  }

\label{sec_discretetree}

In the field of turbulence (see the book \cite{frisch} and references therein),
random cascade models have attracted a lot of interest to describe how the energy injected at the largest scale
flows towards smaller and smaller scales up to the smallest scale where it is dissipated by the viscosity.
The corresponding multifractal properties depend on the choice of the statistics of the cascade generators $W(.)$,
and many different choices have actually been considered in the literature, including 
log-normal \cite{kolmogorov}, bimodal \cite{meneveau}, log-stable \cite{kida,schmitt}, log-Poisson \cite{she_lev,dubrulle,she}, log-infinitely-divisible \cite{novikov}.
Here we will thus consider that the cascade generators $W(.)$ are given, and we will focus on the 
dynamical fluctuations of the empirical dynamics.

Of course besides turbulence, many other applications involve flows on trees with injection at the root,
so that we will keep a general terminology.

\subsection{ Open Markov dynamics in discrete time on the tree }

\label{sec_tree}

We consider a tree of branching $b$, 
starting at the root $(0)$, 
with $b$ sites $(i_1)$ with $i_1=1,2,..,b$ at the first generation $m=1$,
$b^2$ sites $(i_1,i_2)$ at the second generation $m=2$, etc,
up to $b^M$ sites $(i_1,i_2,..,i_M)$ at the last generation $m=M$.

The open Markov dynamics on this tree is directed between the root $(0)$ representing a 'source reservoir',
where the occupation number $N^{ext}_t(0)$ is fixed by the external protocol,
and the $b^M$ sites $(i_1,i_2,..,i_M)$ of the last generation $m=M$ that represent 'sink reservoirs'
that absorb all arriving particles.
The system corresponds to the sites belonging to generations $1 \leq m \leq M-1$.

Between $t$ and $(t+1)$, a particle on the root reservoir $(0)$ at time $t$ can 
 move towards one of the $b$ sites $i_1=1,..,b$ of the first generation $m=1$ with probabilities $W_t(x=i_1,r=0) \equiv W_t(i_1)$ normalized to
\begin{eqnarray}
1 = \sum_{i_1=1}^{b} W_t(i_1) 
\label{normawoutofroot}
\end{eqnarray}

Between $t$ and $(t+1)$, a particle on the system site $(i_1,..,i_m)$ of generation $m$ at time $t$ can 
 move towards one of the $b$ sites $(i_1,..,i_m,i_{m+1})$ of the next generation $(m+1)$ with probabilities $W_t((i_1,i_m,i_{m+1});(i_1,...i_m)) \equiv W_t(i_1,..,i_{m+1})$ normalized to
\begin{eqnarray}
1 = \sum_{i_{m+1}=1}^{b} W_t(i_1,...,i_m,i_{m+1}) 
\label{normawoutoftree}
\end{eqnarray}

\subsection{ Probability of an empirical history of the occupation numbers $N_{t}(.)$ and flows $Q_{t}(.)$ }

The application of Eq. \ref{ptrajmarkovchain} to the open Markov dynamics described above
yields that
the joint probability of the empirical occupation numbers $N_{t}(i_1,..,i_m)$ for the generations belonging to the system $1 \leq m \leq M-1$ and of the
empirical flows $Q_t((i_1,i_m,i_{m+1});(i_1,...i_m)) \equiv Q_t(i_1,..,i_{m+1})$
for $0 \leq m \leq M-1$ reads
\begin{eqnarray}
&& {\cal P} [ N_{1 \leq t \leq T}(.) ; Q_{0 \leq t \leq T-1}(.) \vert N_0(.)] =
e^{ - \displaystyle 
\sum_{t=0}^{T-1}   \sum_{i_1=1}^{b} Q_t(i_1)   \ln \left( \frac{Q_t(i_1)  }{  W_t(i_1) N^{ext}_t(0)} \right)
   }
\prod_{t=0}^{T-1}  \delta_{N^{ext}_{t}(0) , \sum_{i_1 =1}^{b}  Q_t(i_1)  }
\nonumber \\
&&  
e^{ - \displaystyle 
\sum_{t=0}^{T-1} 
\sum_{m=1}^{M-1} \sum_{i_1=1}^b ... \sum_{i_m=1}^b
\sum_{i_{m+1} =1}^{b} Q_t(i_1,..,i_{m+1} )   \ln \left( \frac{Q_t(i_1,..,i_{m+1} )  }{  W_t(i_1,..,i_{m+1}) N_t(i_1,..,i_m)} \right)   }
\nonumber \\
&&  \prod_{t=0}^{T-1} \prod_{m=1}^{M-1} \prod_{i_1=1}^b ... \prod_{i_m=1}^b
\delta_{N_{t}(i_1,..,i_m) , \sum_{i_{m+1} =1}^{b}  Q_t(i_1,..,i_{m+1} )  }
\nonumber \\
&&  \prod_{t=0}^{T-1} \prod_{m=1}^{M-1} \prod_{i_1=1}^b ... \prod_{i_m=1}^b
\delta_{N_{t+1}(i_1,..,i_m) , Q_t(i_1,..,i_m)  }
\label{ptrajmarkovchaintree}
\end{eqnarray}

\subsection{ Probability of an empirical history of the flows $Q_{t}(.)$ alone}

The last line of Eq. \ref{ptrajmarkovchaintree}
simply means that in this directed model, the occupation number $N_{t+1}(i_1,..,i_m) $
of a system site of generation $1 \leq m \leq M-1$ coincides with the incoming flow $Q_t(i_1,..,i_m)  $
from its ancestor. As a consequence, these constraints can be used to eliminate the 
occupation numbers to obtain the probability of an empirical history of the flows $Q_{t}(.)$ alone
\begin{eqnarray}
 {\cal P} [  Q_{0 \leq t \leq T-1}(.) \vert N_0(.)] && =
e^{ - \displaystyle 
\sum_{t=0}^{T-1}  \sum_{i_1=1}^{b} Q_t(i_1)   \ln \left( \frac{Q_t(i_1)  }{  W_t(i_1) N^{ext}_t(0)} \right)
  }
\prod_{t=0}^{T-1}  \delta_{N^{ext}_{t}(0) , \sum_{i_1 =1}^{b}  Q_t(i_1)  }
\nonumber \\
&&  
e^{ - \displaystyle 
\sum_{m=1}^{M-1} \sum_{i_1=1}^b ... \sum_{i_m=1}^b
  \sum_{i_{m+1} =1}^{b} Q_0(i_1,..,i_{m+1} )   \ln \left( \frac{Q_0(i_1,..,i_{m+1} )  }{  W_0(i_1,..,i_{m+1}) N_0(i_1,..,i_m)} \right)   }
\nonumber \\
&&   \prod_{m=1}^{M-1} \prod_{i_1=1}^b ... \prod_{i_m=1}^b
\delta_{N_{0}(i_1,..,i_m) , \sum_{i_{m+1} =1}^{b}  Q_0(i_1,..,i_{m+1} )  }
\nonumber \\
&&  
e^{ - \displaystyle 
\sum_{t=1}^{T-1} 
\sum_{m=1}^{M-1} \sum_{i_1=1}^b ... \sum_{i_m=1}^b
\sum_{i_{m+1} =1}^{b} Q_t(i_1,..,i_{m+1} )   \ln \left( \frac{Q_t(i_1,..,i_{m+1} )  }{  W_t(i_1,..,i_{m+1}) Q_{t-1}(i_1,..,i_m)} \right)  }
\nonumber \\
&&  \prod_{t=1}^{T-1} \prod_{m=1}^{M-1} \prod_{i_1=1}^b ... \prod_{i_m=1}^b
\delta_{Q_{t-1}(i_1,..,i_m) , \sum_{i_{m+1} =1}^{b}  Q_t(i_1,..,i_{m+1} )  }
\label{ptrajmarkovchaintreeflows}
\end{eqnarray}

\subsection{Application to time-independent dynamics during a large time interval $T \to +\infty$ }

\subsubsection{ Large deviations for the flows} 

When the dynamics corresponds to time-independent transition probabilities $W_t(.)=W(.)$
and time-independent reservoir occupation number at the root $  N_t^{ext}(0) = N^{ext}(0)$,
Eq. \ref{ptrajmarkovchaintreeflows} yields that 
the probability to observe some stationary flows $Q_t(.)=Q(.,.)$ during a large time interval $T \to +\infty$
follows the large deviation form with respect to $T$
\begin{eqnarray}
 {\cal P}^{statio} [ Q(.) ] \opsimeq_{T \to +\infty}  && 
e^{ - \displaystyle 
T  \left[ \sum_{i_1=1}^{b} Q(i_1)   \ln \left( \frac{Q(i_1)  }{  W(i_1) N^{ext}(0)} \right)
  \right] }
\nonumber \\
&&  
e^{ - \displaystyle 
T
\sum_{m=1}^{M-1} \sum_{i_1=1}^b ... \sum_{i_m=1}^b
 \left[ \sum_{i_{m+1} =1}^{b} Q(i_1,..,i_{m+1} )   \ln \left( \frac{Q(i_1,..,i_{m+1} )  }{  W(i_1,..,i_{m+1}) Q(i_1,..,i_m)} \right)  \right] }
 \nonumber \\ && 
 \delta_{N^{ext}(0) , \sum_{i_1 =1}^{b}  Q(i_1)  }  \prod_{m=1}^{M-1} \prod_{i_1=1}^b ... \prod_{i_m=1}^b
\delta_{Q(i_1,..,i_m) , \sum_{i_{m+1} =1}^{b}  Q(i_1,..,i_{m+1} )  }
\label{pstatiomarkovchaintreeflows}
\end{eqnarray}

\subsubsection{ Generating function of the flows} 

Equivalently, if one consider the generating function of all the flows $Q(.)$ of the tree
\begin{eqnarray}
Z^{statio} [ \nu(.) ]    
&& \equiv \sum_{Q(.) } P^{statio} [ Q(.)  ]  \ \ e^{ \displaystyle T \sum_{m=1}^M \sum_{i_1=1}^{b} ...\sum_{i_m=1}^{b} 
 \nu(i_1,..,i_m) Q(i_1,..,i_m) }
\label{genetree}
\end{eqnarray}
the explicit expression analog to Eq. \ref{genemultinomialres} for the multinomial distribution at each node
can be used iteratively to obtain the final result
\begin{eqnarray}
Z^{statio}[ \nu(.) ]    
= \left[ \sum_{i_1=1}^{b} W(i_1) e^{\nu(i_1)  } \sum_{i_2=1}^b W(i_1 i_2) e^{\nu(i_1 i_2)  } ...  
\sum_{i_M=1}^b W(i_1 ... i_M) e^{\nu(i_1 ... i_M)  }
\right]^{T N^{ext}(0)}
\label{genetreeres}
\end{eqnarray}

In particular, the joint distribution of the $b^M$ outgoing flows $Q(i_1,..,i_M)$ at the last generation $m=M$ of the tree
has for generating function
\begin{eqnarray}
Z^{statio}[ \nu(i_1,..,i_M) ]  &&   \equiv \sum_{Q(.) } P^{statio} [ Q(.)  ]  \ \ e^{ \displaystyle T  \sum_{i_1=1}^{b} ...\sum_{i_M=1}^{b} 
 \nu(i_1,..,i_M) Q(i_1,..,i_M) }
\nonumber \\
&& = \left[ \sum_{i_1=1}^{b} \sum_{i_2=1}^b  ...  \sum_{i_{M-1}=1}^{b}
\sum_{i_M=1}^b W(i_1)W(i_1 i_2)... W(i_1 ... i_{M-1})  W(i_1 ... i_M) e^{\nu(i_1 ... i_M)  }
\right]^{T N^{ext}(0)}
\label{genetreereslast}
\end{eqnarray}
corresponding to a Bernoulli distribution 
with the $b^M$ parameters given by the strings of probabilities along each branch 
\begin{eqnarray}
{\cal W}( i_1 ... i_M) \equiv W(i_1)W(i_1 i_2)... W(i_1 ... i_{M-1})  W(i_1 ... i_M)
\label{wstring}
\end{eqnarray}
As a consequence, the probability distribution of a single outgoing flow $Q(i_1=1,i_2=1,..,i_M=1)$
at the last generation $m=M$ of the tree
has for generating function
\begin{eqnarray}
Z^{statio}[ \nu ]  &&   \equiv \sum_{Q(.) } P^{statio} [ Q(.)  ]  \ \ e^{ \displaystyle T   \nu Q(i_1=1,i_2=1..,i_M=1) }
\nonumber \\
&& = \left[ 1+  {\cal W} (1,1,..,1,1) (e^{\nu}-1)   \right]^{T N^{ext}(0)}
= e^{T N^{ext}(0) \ln \left[ 1+  {\cal W} (1,1,..,1,1) (e^{\nu}-1)   \right]} 
\label{genetreereslast1}
\end{eqnarray}
will become the generating function of a Poisson variable in the limit of large $M$ where 
$ {\cal W} (1,1,..,1,1) $ is sufficiently small to linearize the logarithm
\begin{eqnarray}
Z^{statio}[ \nu ]  &&   \opsimeq e^{T N^{ext}(0)  {\cal W} (1,1,..,1,1) (e^{\nu}-1) } 
\label{genetreereslast1poisson}
\end{eqnarray}

\subsubsection{ Large deviations for the empirical transition probabilities }

As a final remark, let us mention that, if instead of the extensive flows $Q(.)$, one wishes to consider the empirical transition probabilities (Eq. \ref{p1k})
\begin{eqnarray}
K(i_1) && \equiv \frac{Q(i_1)  }{   N^{ext}(0)} 
\nonumber \\
K(i_1,..,i_{m+1}) && \equiv \frac{Q(i_1,..,i_{m+1} )}{Q(i_1,..,i_m)}
\label{ktree}
\end{eqnarray}
Eq. \ref{pstatiomarkovchaintreeflows} translates into the large deviation form
\begin{eqnarray}
 {\cal P}^{statio} [ K(.) ] \opsimeq_{T \to +\infty}  && 
e^{ - \displaystyle 
T N^{ext}(0)  \left[ \sum_{i_1=1}^{b} K(i_1)   \ln \left( \frac{K(i_1)  }{  W(i_1) } \right)
  \right] }
\nonumber \\
&&  
e^{ - \displaystyle 
T N^{ext}(0)
\sum_{m=1}^{M-1} \sum_{i_1=1}^b ... \sum_{i_m=1}^b
 \left[ \sum_{i_{m+1} =1}^{b} K(i_1) K(i_1,i_2) ...K(i_1,..,i_{m+1} )   \ln \left( \frac{K (i_1,..,i_{m+1} )  }{  W(i_1,..,i_{m+1}) } \right)  \right] }
 \nonumber \\ && 
 \prod_{m=1}^{M} \prod_{i_1=1}^b ... \prod_{i_m=1}^b
\delta_{1 , \sum_{i_m =1}^{b}  K(i_1,..,i_{m} )  }
\label{pstatiomarkovchaintreeK}
\end{eqnarray}

\subsection{Application to periodic dynamics   }

It is interesting to write Eq. \ref{periodicmarkovchain} for the random cascade on the tree :
one obtains that the probability distribution of the empirical flows $Q_e(.)$ at even times and
$Q_o(.)$ at odd times follows the large deviation form
\begin{eqnarray}
&& {\cal P}^{periodic} [ Q_e(.) ,Q_o(.)] \opsimeq_{T \to +\infty}   
\label{pperiodicmarkovchaintreeflows}
\\ && 
e^{ - \displaystyle 
\frac{T}{2}  \left[ \sum_{i_1=1}^{b} Q_e(i_1)   \ln \left( \frac{Q_e(i_1)  }{  W_e(i_1) N^{ext}_e(0)} \right)  \right] }
\delta_{N_e^{ext}(0) , \sum_{i_1 =1}^{b}  Q_e(i_1)  } 
\nonumber  \\  &&  
e^{ - \displaystyle 
\frac{T}{2}  \left[ \sum_{i_1=1}^{b} Q_o(i_1)   \ln \left( \frac{Q_o(i_1)  }{  W_o(i_1) N^{ext}_o(0)} \right)  \right] }
\delta_{N_o^{ext}(0) , \sum_{i_1 =1}^{b}  Q_o(i_1)  } 
\nonumber \\  &&  
e^{ - \displaystyle   \frac{T}{2} 
\sum_{m=1}^{M-1} \sum_{i_1=1}^b ... \sum_{i_m=1}^b
 \left[ \sum_{i_{m+1} =1}^{b} Q_e(i_1,..,i_{m+1} )   \ln \left( \frac{Q_e(i_1,..,i_{m+1} )  }{  W_e(i_1,..,i_{m+1}) Q_o(i_1,..,i_m)} \right)  \right] }
\delta_{Q_o(i_1,..,i_m) , \sum_{i_{m+1} =1}^{b}  Q_e(i_1,..,i_{m+1} )  }
\nonumber \\  &&  
e^{ - \displaystyle   \frac{T}{2} 
\sum_{m=1}^{M-1} \sum_{i_1=1}^b ... \sum_{i_m=1}^b
 \left[ \sum_{i_{m+1} =1}^{b} Q_o(i_1,..,i_{m+1} )   \ln \left( \frac{Q_o(i_1,..,i_{m+1} )  }{  W_o(i_1,..,i_{m+1}) Q_e(i_1,..,i_m)} \right)  \right] }
\delta_{Q_e(i_1,..,i_m) , \sum_{i_{m+1} =1}^{b}  Q_o(i_1,..,i_{m+1} )  }
\nonumber
\end{eqnarray}
which is thus very similar to Eq. \ref{pstatiomarkovchaintreeflows} with the additional decomposition into even and odd contributions.

Equivalently, the analog of the generating function of Eq. \ref{genetree}
of all the flows $Q_e(.)$ and $Q_o(.)$ of the tree
\begin{eqnarray}
&& Z^{periodic} [ \nu_e(.), \nu_o(.) ]    
 \equiv 
\nonumber \\
&& \sum_{Q_e(.) ,Q_o(.) }  {\cal P}^{periodic} [ Q_e(.) ,Q_o(.)]   \ \ 
e^{ \displaystyle  \frac{T}{2}  \sum_{m=1}^M \sum_{i_1=1}^{b} ...\sum_{i_m=1}^{b} 
 \left( \nu_e(i_1,..,i_m) Q_e(i_1,..,i_m) + \nu_o(i_1,..,i_m) Q_o(i_1,..,i_m)  \right)  }
\label{genetreeperiodic}
\end{eqnarray}
can be written in the following form generalizing Eq. \ref{genetreeres}, assuming that $M$ is even for definiteness
\begin{eqnarray}
&& Z^{periodic} [ \nu_e(.), \nu_o(.) ]    = 
\nonumber \\
&& \left[ \sum_{i_1=1}^{b} W_e(i_1) e^{\nu_e(i_1)  } \sum_{i_2=1}^b W_o(i_1 i_2) e^{\nu_o(i_1 i_2)  } ...  
 \sum_{i_{M-1}=1}^b W_e(i_1 ... i_{M-1}) e^{\nu_e(i_1 ... i_{M-1}) }
\sum_{i_{M}=1}^b W_o(i_1 ... i_M) e^{\nu_o(i_1 ... i_M)  } \right]^{ \frac{T}{2} N_e^{ext}(0)}
\nonumber \\
&& \left[ \sum_{i_1=1}^{b} W_o(i_1) e^{\nu_o(i_1)  } \sum_{i_2=1}^b W_e(i_1 i_2) e^{\nu_e(i_1 i_2)  } ...  
 \sum_{i_{M-1}=1}^b W_o(i_1 ... i_{M-1}) e^{\nu_o(i_1 ... i_{M-1}) }
\sum_{i_{M}=1}^b W_e(i_1 ... i_M) e^{\nu_e(i_1 ... i_M)  } \right]^{ \frac{T}{2} N_o^{ext}(0)}
\label{genetreeresperiodic}
\end{eqnarray}

\section{ Application to continuous time Markovian cascades on trees  }

\label{sec_continuoustree}

As a comparison to the discrete-time cascade model analyzed in the previous section,
it is now interesting to consider the analogous continuous-time cascade model.

\subsection{ Open Markov dynamics in continuous time on the tree }

We consider the same tree structure as in \ref{sec_tree},
but the directed dynamics from the root towards the leaves is now defined in continuous time with transition rates 
$w_t((i_1,..,i_m),(i_1,..,i_{m-1}))  \equiv w_t(i_1,..,i_m)$ per unit time as in section \ref{sec_continuous}.

\subsection{ Probability of an empirical history of the occupation numbers $N_{t}(.)$ and flows $q_{t}(.,.)$ }

The application of Eq. \ref{ptrajmarkovjump} 
to the directed dynamics on the tree
yields that the probability to observe the empirical occupation numbers $N_{0 \leq t \leq T}(.)$
and the empirical flows $ q_{0 \leq t \leq T}(.,.) $ reads
\begin{eqnarray}
&&  {\cal P} [ N_{0 \leq t \leq T}(.) ; q_{0 \leq t \leq T}(.,.) \vert N_0(.)]
=
 e^{ - \displaystyle 
\int_{0}^{T}dt   \sum_{i_1=1 }^{b} \left[ q_t(i_1)   \ln \left( \frac{q_t(i_1)  }{  w_t(i_1) N_t^{ext}(0)} \right)- q_t(i_1)
+w_t(i_1) N_t^{ext}(0) \right] }
\nonumber \\
&&  
e^{ - \displaystyle 
\int_{0}^{T}dt 
\sum_{m=2}^{M} \sum_{i_1=1}^b ... \sum_{i_m=1}^b 
\left[ q_t(i_1,..,i_{m})   \ln \left( \frac{q_t(i_1,..,i_{m})  }{  w_t(i_1,..,i_{m}) N_t(i_1,...,i_{m-1})} \right)- q_t(i_1,..,i_{m})+w_t(i_1,..,i_{m})  N_t(i_1,...,i_{m-1}) \right] }
\label{ptrajmarkovjumptree}
\end{eqnarray}
provided the empirical dynamics of Eq. \ref{empiricaldynaq} is satisfied on the time interval $t \in [0,T]$ 
for all sites $(i_1,..,i_m)$ of generations $1 \leq m \leq M-1$ of the system
\begin{eqnarray}
\frac{d N_{t}(i_1,..,i_m)}{dt}  
=  q_t(i_1,..,i_{m}) - \sum_{i_{m+1}=1}^b q_t(i_1,..,i_{m+1})
\label{empiricaldynatree}
\end{eqnarray}

\subsection{Application to time-independent dynamics during a large time interval $T \to +\infty$ }

\subsubsection{ Large deviations for occupations and flows } 

When the dynamics corresponds to time-independent transition rates $w_t(.)=w(.)$
and time-independent reservoir occupation number at the root $  N_t^{ext}(0) = N^{ext}(0)$,
Eq. \ref{ptrajmarkovjumptree} and \ref{empiricaldynatree}
yield that 
the probability to observe some stationary occupation numbers $N_t(.)=N(.)$ and
flows per unit time $q_t(.)=q(.,.)$ during a large time interval $T \to +\infty$
follows the large deviation form with respect to $T$
\begin{eqnarray}
&&  {\cal P}^{statio} [ N(.) ; q(.,.) ]
=
 e^{ - \displaystyle 
T   \sum_{i_1=1 }^{b} \left[ q(i_1)   \ln \left( \frac{q(i_1)  }{  w(i_1) N^{ext}(0)} \right)- q(i_1)
+w(i_1) N^{ext}(0) \right] }
\nonumber \\
&&  
e^{ - \displaystyle   T
\sum_{m=2}^{M} \sum_{i_1=1}^b ... \sum_{i_m=1}^b 
\left[ q(i_1,..,i_{m})   \ln \left( \frac{q(i_1,..,i_{m})  }{  w(i_1,..,i_{m}) N(i_1,...,i_{m-1})} \right)- q(i_1,..,i_{m})+w(i_1,..,i_{m})  N(i_1,...,i_{m-1}) \right] }
\nonumber \\
&&  \prod_{m=1}^{M-1} \prod_{i_1=1}^b ... \prod_{i_m=1}^b \delta \left( q(i_1,..,i_{m}) - \sum_{i_{m+1}=1}^b q(i_1,..,i_{m+1})  \right)
\label{pstatiomarkovjumptree}
\end{eqnarray}

If instead of the extensive flows per unit time $q(.)$ one wishes to consider the empirical transition rates
\begin{eqnarray}
k(i_1) && = \frac{q(i_1)  }{   N^{ext}(0)} 
\nonumber \\
k(i_1,..,i_{m+1}) && \equiv \frac{q(i_1,..,i_{m+1} )}{N(i_1,..,i_m)}
\label{kratetree}
\end{eqnarray}
Eq. \ref{pstatiomarkovjumptree} translates into the large deviation form
\begin{eqnarray}
&&  {\cal P}^{statio} [ N(.) ; k(.,.) ]
=
 e^{ - \displaystyle 
T N^{ext}(0)   \sum_{i_1=1 }^{b} \left[ k(i_1)   \ln \left( \frac{k(i_1)  }{  w(i_1) } \right)- k(i_1)
+w(i_1)  \right] }
\nonumber \\
&&  
e^{ - \displaystyle   T 
\sum_{m=2}^{M} \sum_{i_1=1}^b ... \sum_{i_m=1}^b 
N(i_1,...,i_{m-1}) \left[ k(i_1,..,i_{m})   \ln \left( \frac{k(i_1,..,i_{m})  }{  w(i_1,..,i_{m}) } \right)- k(i_1,..,i_{m})+w(i_1,..,i_{m})   \right] }
\nonumber \\
&&  \prod_{m=1}^{M-1} \prod_{i_1=1}^b ... \prod_{i_m=1}^b
 \delta \left( k(i_1,..,i_{m}) N(i_1,...,i_{m-1}) - \sum_{i_{m+1}=1}^b k(i_1,..,i_{m+1}) N(i_1,...,i_{m}) \right)
\label{pstatiomarkovjumptreerates}
\end{eqnarray}
but since the empirical dynamics constraints of the last line involve both $N(.)$ and $k(.)$,
it is more convenient to work with Eq. \ref{pstatiomarkovjumptree} where the empirical dynamics constraints of the last line involve only the flows.

\subsubsection{ Large deviations for the flows alone} 

To obtain the large deviation properties of the flows per unit time $q(.)$ alone,
one needs to optimize Eq. \ref{pstatiomarkovjumptree} over the occupation numbers $N(.)$ :
plugging the optimal values
\begin{eqnarray}
N^{opt}(i_1,..,i_{m-1}) = \frac{ \sum_{j_m=1}^b q(i_1,..,j_{m}) }{ \sum_{j_m=1}^b w(i_1,..,j_{m})}
\label{nopt}
\end{eqnarray}
into Eq. \ref{pstatiomarkovjumptree}
yields that the probability to observe the stationary flows per unit time $q(.)$ reads
\begin{eqnarray}
&&  {\cal P}^{statio} [  q(.,.) ]
=
 e^{ - \displaystyle 
T   \sum_{i_1=1 }^{b} \left[ q(i_1)   \ln \left( \frac{q(i_1)  }{  w(i_1) N^{ext}(0)} \right)- q(i_1)
+w(i_1) N^{ext}(0) \right] }
\nonumber \\
&&  
e^{ - \displaystyle   T
\sum_{m=2}^{M} \sum_{i_1=1}^b ... \sum_{i_m=1}^b 
\left[ q(i_1,..,i_{m})   \ln \left( \frac{q(i_1,..,i_{m}) \sum_{j_m=1}^b w(i_1,..,j_{m}) }{  w(i_1,..,i_{m}) \sum_{j_m=1}^b q(i_1,..,j_{m})} \right)- q(i_1,..,i_{m})+w(i_1,..,i_{m})  
\frac{ \sum_{j_m=1}^b q(i_1,..,j_{m}) }{ \sum_{j_m=1}^b w(i_1,..,j_{m})} \right] }
\nonumber \\
&&  \prod_{m=1}^{M-1} \prod_{i_1=1}^b ... \prod_{i_m=1}^b \delta \left( q(i_1,..,i_{m}) - \sum_{i_{m+1}=1}^b q(i_1,..,i_{m+1})  \right)
\label{pstatiomarkovjumptreeflows}
\end{eqnarray}

\subsubsection{ Large deviations for the occupations alone} 

To obtain the large deviation properties of the occupation numbers $N(.)$ alone,
one needs to optimize Eq. \ref{pstatiomarkovjumptree} over the flows $q(.)$ :
the iterative optimization starting from the last generation yields that
the probability to observe the stationary occupations numbers $N(.)$ takes the following form
\begin{eqnarray}
&&  {\cal P}^{statio} [  N(.) ]
=
 e^{ - \displaystyle 
T \left[ \sum_{m=1}^{M} \sum_{i_1=1}^b ... \sum_{i_m=1}^b \lambda_{i_1,...,i_m} 
- M \sum_{i_1=1}^b c(i_1)
\right] }
\label{pstatiomarkovjumptreeoccup}
\end{eqnarray}
in terms of the typical flows associated to the occupation numbers $N(.)$
\begin{eqnarray}
\lambda_{i_1,...,i_m} \equiv  N(i_1,...,i_{m-1}) w(i_1,..,i_{m})
\label{notalambda}
\end{eqnarray}
and of the numbers $c(i_1)$ that should be computed by the following recurrence starting at the last generation $m=M$
\begin{eqnarray}
c_{i_1,...,i_M} && \equiv \lambda_{i_1,...,i_M} 
\nonumber \\
c_{i_1,...,i_{M-1}} && \equiv
\left[ \lambda_{i_1,...,i_{M-1}} \left( \sum_{i_M=1}^b c_{i_1,...,i_M} \right) \right]^{\frac{1}{2}}
\nonumber \\
c_{i_1,...,i_{M-2}} && \equiv 
\left[ \lambda_{i_1,...,i_{M-2}} \left( \sum_{i_{M-1}=1}^b c_{i_1,...,i_{M-1}} \right)^2 \right]^{\frac{1}{3}}
\nonumber \\
..................... && ..................................................................................
\nonumber \\
c_{i_1,i_2} && \equiv 
\left[ \lambda_{i_1,i_2} \left( \sum_{i_3=1}^b c_{i_1,i_2,i_3} \right)^{M-2} \right]^{\frac{1}{M-1}}
\nonumber \\
c_{i_1} && \equiv 
\left[ \lambda_{i_1} \left( \sum_{i_2=1}^b c_{i_1,i_2} \right)^{M-1} \right]^{\frac{1}{M}}
\label{reccurrencec}
\end{eqnarray}
Note that the typical occupation numbers 
\begin{eqnarray}
N^{typ}(i_1,...,i_m )= N^{ext}(0) \left( \frac{ w_{i_1} }{\sum_{j_2} w_{i_1,j_2} } \right) \left(  \frac{ w_{i_1} }{\sum_{j_2} w_{i_1,j_2} } \right)
... \left( \frac{ w_{i_1,...,i_m} }{\sum_{j_{m+1}} w_{i_1,...,j_{m+1}} } \right)
\label{ntyptree}
\end{eqnarray}
where the large deviation rate function of Eq. \ref{pstatiomarkovjumptreeoccup} vanishes,
corresponds to the case where the associated $\lambda^{typ}_{i_1,...,i_m}  $ of Eq. \ref{notalambda}
satisfy the sum rules
\begin{eqnarray}
\sum_{i_m} \lambda_{i_1,...,i_m}^{typ} = \lambda_{i_1,...,i_{m-1}}^{typ}
\label{sumrulelambda}
\end{eqnarray}
so that the corresponding numbers $c^{typ}(i_1,..,i_m)$ of the recurrence of Eq. \ref{reccurrencec}
then coincide with the $\lambda^{typ}(i_1,..,i_m)$.
For non-typical occupation numbers $N(.) \ne N^{typ}(.)$, 
the rate function of Eq. \ref{pstatiomarkovjumptreeoccup}
is not completely explicit in terms of the $N(.)$ since one should first solve the recurrence of Eq \ref{reccurrencec}
on the tree.

\subsection{ Special case $b=1$ for the one-dimensional random directed model between a source and a sink }

For the special case of branching ratio $b=1$, the tree reduces to the one-dimensional lattice of $(M-1)$ sites
labelled by the generation $m=1,2,..,M-1$ [instead of the previous tree notation $(i_1=1,i_2=1,..,i_m=1)$]
characterized by occupation numbers $N(m)$ and incoming flow $q(m,m-1)$
between the 'source reservoir' at $m=0$ and the 'sink reservoir' at $m=M$.
 The constraints of the empirical dynamics written on the last line of Eq. \ref{pstatiomarkovjumptree}
impose that all these flow takes the same value $j$
\begin{eqnarray}
j= q(1,0) = q(2,1) = .. = q(M,M-1) =j
\label{j1d}
\end{eqnarray}
So Eq. \ref{pstatiomarkovjumptree}
yields that 
the probability to observe some stationary occupation numbers $N(.)$ and
the current $j$
follows the large deviation form with respect to $T$
\begin{eqnarray}
  {\cal P}^{statio} [ N(.) ;j ]
=
&& 
 e^{ - \displaystyle 
T  \left[  j  \ln \left( \frac{ j  }{  w(1,0) N^{ext}(0)} \right)- j +w(1,0) N^{ext}(0) \right] }
\nonumber \\
&&  
e^{ - \displaystyle   T
\sum_{m=1}^{M-1} 
\left[ j  \ln \left( \frac{j }{  w(m+1,m) N(m)} \right)- j+w(m+1,m) N(m)  \right] }
\label{pstatiomarkovjumptreeb1}
\end{eqnarray}
that can be compared with the rate function of Eq (36) in Ref. \cite{c_ring} 
concerning the rate function of the same directed one-dimensional model 
defined on a ring geometry
that conserves the number of particles, with the 
correspondence $w(m+1,m) = \frac{1}{\tau_m}$ with the random trapping times $\tau_m$ used in Ref. \cite{c_ring} .

The optimization of Eq. \ref{pstatiomarkovjumptreeb1} with respect to the occupation numbers $N(.)$ yields
\begin{eqnarray}
N^{opt}(m) = \frac{j}{ w(m+1,m)} 
\label{nopt1d}
\end{eqnarray}
so that the probability of the current $j$ alone is reduced to the first term involving the source
\begin{eqnarray}
  {\cal P}^{statio} [ j ]
= e^{ - \displaystyle 
T  \left[  j  \ln \left( \frac{ j  }{  w(1,0) N^{ext}(0)} \right)- j +w(1,0) N^{ext}(0) \right] }
\label{pstatiomarkovjumptreeb1jalone}
\end{eqnarray}
and is thus completely different from the corresponding result for the ring geometry \cite{c_ring} 
where the conservation of the total number of particles requires the introduction of a Lagrange multiplier in the optimization.

The optimization of Eq. \ref{pstatiomarkovjumptreeb1} with respect to the current $j$ yields
\begin{eqnarray}
j^{opt} = \left( w(1,0) N^{ext}(0) \prod_{m=1}^{M-1} w(m+1,m) N(m) \right)^{\frac{1}{M}}
\label{jopt1d}
\end{eqnarray}
so that the probability of the occupation numbers $N(.)$  
\begin{eqnarray}
  {\cal P}^{statio} [ N(.)  ]
=
 e^{ - \displaystyle 
T  \left[ w(1,0) N^{ext}(0)+\sum_{m=1}^{M-1} w(m+1,m) N(m)  - M 
\left( w(1,0) N^{ext}(0)\prod_{m=1}^{M-1} w(m+1,m) N(m) \right)^{\frac{1}{M}}  \right] }
\label{pstatiomarkovjumptreeb1nalone}
\end{eqnarray}
is very similar to the result found for ring geometry \cite{c_ring}.

More details on the large deviation properties of the random trap model on the ring
can be found in Refs \cite{c_ring,vanwijland_trap} (see also \cite{BFG2} where it is called the 'random watch' model), since here our purpose was only to stress the differences
induced by the presence of reservoirs.


\section{ Conclusion }

\label{sec_conclusion}

In this paper, we have extended the large deviations at 'Level 2.5 in time' for time-dependent ensemble-empirical-observables, introduced in Ref \cite{maes_onandbeyond} for the case of a fixed number $N$ of independent Markov jump processes, to the case of open Markov processes with reservoirs : we have derived explicit formulas for the joint probability of empirical occupation numbers and empirical flows, both for discrete-time dynamics and for continuous-time jump dynamics, with possibly time-dependent dynamical rules and/or time-dependent driving of the reservoirs. We have then applied this general formalism to random cascades on trees, where particles are injected at the root via a 'source reservoir', while the particles are removed at the leaves of the last generation of the tree via 'sink reservoirs', again both for discrete-time dynamics and for continuous-time jump dynamics. Finally, we have also mentioned the results for the one-dimensional trap model between a source and a sink, in order to compare with the large deviations obtained
for the same model defined on a ring geometry where the total number of particles is conserved \cite{c_ring}. 
In the future, it would be thus interesting to apply this general formalism to other interesting open Markov processes.

As a final remark, let us mention that the present approach has been generalized further to the presence of interactions 
 in the recent preprint \cite{c_ldinter}.

\appendix

\section{Reminder on the 'Level 2.5' for the time-empirical-observables of a single Markov process }

\label{app_reminder}

In this Appendix, we briefly recall the Large deviations 'at Level 2.5' for the time-empirical-observables of a single Markov chain
and a single Markov jump process respectively, in order to make comparisons with equations of the text.

\subsection{ Case of a single Markov Chain \cite{fortelle_thesis,fortelle_chain,review_touchette} }

\label{app_singlemarkovchain}

The probability $P_t(x)$ to be in configuration $x$ at time $t$ evolves according to
the discrete-time Markov Chain
\begin{eqnarray}
P_{t+1}(x) = \sum_y W(x,y) P_t(y) 
\label{markov}
\end{eqnarray}
where the transition probabilities $W(x,y) \geq 0 $ from $y$ to $x$ satisfy the normalization for each $y$
\begin{eqnarray}
 \sum_x  W(x,y)=1
\label{markovnormaw}
\end{eqnarray}

If one starts at time $t=0$ with some initial distribution state $P_{t=0}(x_0)$,
the probability of the whole  trajectory $(x_0,x_1,x_2,..,x_T)$ 
\begin{eqnarray}
{\cal P}[ x_{0 \leq t \leq T} ]= W(x_T,x_{T-1}) ....   W(x_2,x_1) W(x_1,x_0) P_{0}(x_0)
= e^{ \displaystyle \sum_{t=1}^T  \ln W(x_t,x_{t-1})  } P_{0}(x_0) 
\label{ptrajchain}
\end{eqnarray}
can be rewritten in terms of the time-empirical flows
\begin{eqnarray}
Q^{te} (y, x) \equiv \frac{1}{T}  \sum_{t=0}^{T-1} \delta_{x(t+1),y} \delta_{x(t),x} 
\label{empiricalmatrix}
\end{eqnarray}
as
\begin{eqnarray}
{\cal P}[ x_{0 \leq t \leq T} ]
= e^{ \displaystyle T \sum_{x} \sum_y Q^{te} (x, y)   \ln W(x,y)  } P_{0}(x_0) 
\label{ptrajchainte}
\end{eqnarray}

From $Q^{te} (y, x)  $ introduced in Eq. \ref{empiricalmatrix},
the time-empirical density $ \rho^{te}(x)  $ can be reconstructed via the sum over the final point $y$ 
\begin{eqnarray}
 \rho^{te}(x) \equiv  \frac{1}{T}  \sum_{t=0}^{T-1} \delta_{x(t),x} = \sum_y Q^{te} (y, x)
\label{empiricaldensity}
\end{eqnarray}
or via the sum over the initial point $x$, up to boundary terms that become negligible in the limit of large time $T \to +\infty$
\begin{eqnarray}
\sum_x Q^{te} (y, x) = \frac{1}{T}  \sum_{t=0}^{T-1} \delta_{x(t+1),y} =  \rho^{te}(y) +\frac{ \delta_{x(T),y} - \delta_{x(0),y}  }{T}
\label{empiricalmatrixb}
\end{eqnarray}

Here the 'Level 2.5' statement \cite{fortelle_thesis,fortelle_chain,review_touchette}
is that the probability to observe 
the time-empirical flows $Q^{te}(.,.)$ follows the Large Deviation Form with respect to the large time $T \to +\infty$
\begin{eqnarray}
P_T[ Q^{te}(.,. ) ] \oppropto_{T \to +\infty} e^{- T \displaystyle \sum_x  \sum_y Q^{te}(y,x) \ln \left( \frac{  Q^{te}(y,x) }{ W(y,x) \rho^{te}(x)   }  \right) }
\label{alevel2.5chain}
\end{eqnarray}
where $Q^{te}(x,y)$ satisfies the stationarity constraint in relation with the time-empirical density $  \rho^{te}(x) $
\begin{eqnarray}
 \sum_y Q^{te}(x, y)  = \sum_y Q^{te}(y,x)  = \rho^{te}(x) 
\label{Cstatio}
\end{eqnarray}
Eq. \ref{alevel2.5chain} can be translated for the empirical transition probabilities 
\begin{eqnarray}
K^{te}(y,x) = \frac{Q^{te} (y, x) }{  \rho^{te}(x)  } 
\label{empiricalK}
\end{eqnarray}
into
\begin{eqnarray}
P_T[ K^{te}(.,. ) ] \oppropto_{T \to +\infty} e^{- T \displaystyle \sum_x  \sum_y K^{te}(y,x) \rho^{te}(x) \ln \left( \frac{  K^{te}(y,x) }{ W(y,x)    }  \right) }
\label{alevel2.5chainK}
\end{eqnarray}
where Eq. \ref{Cstatio} yields the constraints
\begin{eqnarray}
\sum_y K^{te}(y,x)  && =1
\nonumber \\
 \sum_y K^{te}(x, y)  \rho^{te}(y)   && = \rho^{te}(x) 
\label{CstatioK}
\end{eqnarray}

\subsection{ Case of 
a single Markov Jump process \cite{fortelle_thesis,fortelle_jump,wynants_thesis,maes_canonical,maes_onandbeyond,chetrite_formal,BFG1,BFG2}}

\label{app_singlemarkovjump}

The probability $P_t(x)$ to be in configuration $x$ at time $t$ evolves according to
the continuous-time
Master Equation
\begin{eqnarray}
\partial_t P_{t}(x) = \sum_{y \ne x} w(x,y)  P_t(y)- \sum_{y \ne x} w(y,x)   P_t(x)
\label{master}
\end{eqnarray}

The probability of the whole trajectory $(x_t)_{0 \leq t \leq T}$
\begin{eqnarray}
 {\cal P} \left( [x(t)]_{0 \leq t \leq T} \right) 
=  e^{ \displaystyle \sum_{ t: x(t^-) \ne x(t^+)  }  \ln w(x(t^+) ,x(t^-) )  - \int_0^T dt  \sum_{y \ne x} w(y,x(t) ) } P_{0}(x_0)
\label{ptraject}
\end{eqnarray}
can be rewritten in terms of the time-empirical density 
\begin{eqnarray}
\rho^{te}(x) \equiv  \frac{1}{T} \int_0^T dt \delta_{x(t),x}
\label{empiricaldensitycontinuous}
\end{eqnarray}
and of the time-empirical jump density 
\begin{eqnarray}
q^{te}(y ,x) \equiv  \frac{1}{T} \sum_{t : x(t) \ne x(t^+)} \delta_{x(t^+),y} \delta_{x(t),x} 
\label{jumpempiricaldensity}
\end{eqnarray}
as
\begin{eqnarray}
 {\cal P} \left( [x(t)]_{0 \leq t \leq T} \right) 
=  e^{ \displaystyle T\sum_{x } \sum_{y \ne x}\left[ q^{te}(y ,x) \ln w(y,x )  -  w(y,x) \rho^{te}(x) \right] } P_{0}(x_0)
\label{ptrajecte}
\end{eqnarray}

Here the 'Level 2.5' statement \cite{fortelle_thesis,fortelle_jump,wynants_thesis,maes_canonical,maes_onandbeyond,chetrite_formal,BFG1,BFG2}
is that the probability to observe 
the time-empirical density $\rho^{te}(.)$ 
and the time-empirical jump density $q^{te}(.,.) $
follows the Large Deviation form with respect to the large time $T \to +\infty$
\begin{eqnarray}
P_T[ \rho(.) ; q(. , .) ]
 \oppropto_{T \to +\infty} e^{- \displaystyle T  \sum_x  \sum_{y \ne x}
\left[ q^{te}(y,x)  \ln \left( \frac{ q^{te}(y,x)  }{  w(y,x) \rho^{te}(x) }  \right) 
 - q^{te}(y,x)  + w(y,x) \rho^{te}(x)  \right] }
\label{level2.75master}
\end{eqnarray}
where $q^{te}(y,x)  $ 
 should satisfy the stationarity constraint
\begin{eqnarray}
\sum_{y \ne x} q^{te}(y,x)  = \sum_{y \ne x} q^{te}(x,y)  
\label{contrainteq}
\end{eqnarray}

In terms of the time-empirical transition rates
\begin{eqnarray}
k^{te}(y,x) = \frac{q^{te} (y, x) }{  \rho^{te}(x)  } 
\label{empiricalk}
\end{eqnarray}
Eq. \ref{level2.75master} can be translated into
 \begin{eqnarray}
P_T[ \rho(.) ; k(. , .) ]
 \oppropto_{T \to +\infty} e^{- \displaystyle T  \sum_x  \sum_{y \ne x}
\left[ k^{te}(y,x)  \ln \left( \frac{ k^{te}(y,x)  }{  w(y,x)  }  \right) 
 - k^{te}(y,x)  + w(y,x)   \right]\rho^{te}(x)   }
\label{level2.75masterk}
\end{eqnarray}
where the stationarity constraint of Eq. \ref{contrainteq} becomes
\begin{eqnarray}
\sum_{y \ne x} k^{te}(y,x) \rho^{te}(x)  = \sum_{y \ne x} k^{te}(x,y)  \rho^{te}(y) 
\label{contraintek}
\end{eqnarray}


\end{document}